\documentclass[aps,prl,preprint,superscriptaddress,showpacs,showkeys]{revtex4}

\usepackage{graphics}
\usepackage{longtable}
\begin{document}

\title{Native geometry and the dynamics of protein folding}

\author{P. F. N. Faisca and M. M. Telo da Gama}
\email{patnev@alf1.cii.fc.ul.pt}
\affiliation{CFTC, Av. Prof. Gama Pinto 2, 1649-003 Lisboa Codex, Portugal}

\date{\today}

\begin{abstract}
In this paper we investigate the role of native geometry on the kinetics        
of protein folding based on simple lattice models and Monte Carlo               
simulations.                                                                    
Results obtained within the scope of the Miyazawa-Jernigan indicate the         
existence of two dynamical folding regimes depending on the protein chain       
length. For chains larger than 80 amino acids the folding performance is        
sensitive to the native state's conformation.                                   
Smaller chains, with less than 80 amino acids, fold via two-state               
kinetics and exhibit a significant correlation between the contact order        
parameter and the logarithmic folding times. In particular, chains              
with N=48 amino acids were found to belong to two broad classes of              
folding, characterized by different cooperativity, depending on the contact     
order parameter.                                                                
Preliminary results based on the G\={o} model show that the effect of long      
range contact interaction strength in the folding kinetics is largely           
dependent on the native state's geometry. 
\end{abstract}
\maketitle

\section{Introduction}
It is well known that most small (from $\sim$ 50-120 amino acids),
single domain proteins fold via a two-state (single exponential) kinetics, 
without observable folding intermediates and with a single transition state 
associated with one major free energy barrier separating the native state 
from the unfolded conformations~\cite{JACKSON,PLAXCO0,KAYA0}. For this reason 
small protein molecules are particularly well suited for investigating 
the correlations between folding times and the native
state equilibrium properties, a major challenge for those working in protein 
research.\par
The energy landscape theory predicts that the landscape's rugedeness
plays a fundamental role in the folding kinetics of model proteins: 
The existence of local energy minima, that act as kinetic traps, is 
responsible for the overall slow and, under some conditions (as the 
temperature is lowered towards the glass transition temperature), glassy 
dynamics~\cite{WOLYNES}. On the other hand, rapid folding is associated 
with the existence of a smooth, funnel-shaped energy landscape~\cite{ONUCHIC}. 
In Refs.~\cite{PLAXCO1,GILLESPIE} Plaxco {\it et. al.} and
Gillespie and Plaxco have provided experimental evidence
that the folding energy landscape of single domain proteins is extremly 
smooth even at considerably low temperatures. Therefore 
differences in the landscape's `topography'
cannot account for the vast range of folding rates as observed in real 
proteins
~\cite{LEE, DOBSON}. However, 
a strong correlation (r=0.94) was found between the so-called contact order 
parameter, CO, and the experimentally observed folding rates in a set of 24 
non-homologous single domain proteins~\cite{PLAXCO2}. The CO measures
the average sequence separation of contacting residue pairs
in the native structure relative to the chain length of the protein
\begin{equation}
CO=\frac{1}{LN}\sum_{i,j}^N \Delta_{i,j}\vert i-j \vert,
\label{eq:no1}
\end{equation}  
where $\Delta_{i,j}=1$ if residues $i$ and $j$ are in contact and is 0 
otherwise; $N$ is the total number of contacts and $L$ is the protein chain 
length.
The empirical observation that the CO correlates well with the folding rates 
of single domain proteins, exhibiting smooth energy landscapes, strongly 
suggests a geometry-dependent kinetics for such two-state folders. \par
The connection between the CO and the dominant range of residue interactions
brings back an old, well-debated issue in the protein folding
literature, that of the role of local (i.e. close in space and in sequence) 
and long range 
(i.e. close in space but distant along the sequence)
inter-residue interactions in the folding dynamics. Several results appear
to agree on the idea that long range (LR) contacts play an active role in 
stabilizing the native
fold {~\cite{GO, GUTIN, BAKER, GROMIHA0}}. 
In what regards the folding 
kinetics, results reported in Refs.~\cite{GO, BAKER, WETLAUFER, MOULT}
suggest that local contacts increase the folding speed, relative to LR 
contacts, while results in Ref.~\cite{GUTIN} suggest an opposite trend. 
In Ref.~\cite{GROMIHA2} Gromiha and Selvaraj have analysed explicitly 
the contribution of LR contacts in determining the folding rates 
of 23 (out of the 24) two-state folders studied by 
Plaxco {\it et al}~\cite{PLAXCO2}.
These authors proposed the so-called long range order (LRO) parameter, 
measuring the total number of long range contacts relative to the protein 
chain length, as an alternative way of quantifying the native structure 
geometry. In fact, the LRO parameter correlates as well as the CO with the 
folding rates of the two-state folders 
analysed in Ref.~\cite{PLAXCO2}.\par
The majority of protein folding theory is based,  
not only on results for real proteins such as those outlined above, but 
also 
on a vast number of findings obtained within the scope of simple lattice 
models and Monte Carlo (MC) simulations. Although lattice models do not 
encompass the full complexity of real proteins they
are non trivial and capture fundamental aspects of the protein 
folding kinetics~\cite{SFERMI}.\par
In the present study we investigate through Monte Carlo folding simulations
of simple lattice models, such as the Miyazawa-Jernigan model and the 
G\={o} model, the 
dependence of two-state folding kinetics on the native state geometry. \par

\section{LATTICE MODELS \label{sec:secno2}}

In a lattice model the protein is reduced to its backbone structure:
amino acids are represented by beads of uniform size, occupying the
lattice vertices, and the peptide bond, that covalently connects amino
acids along the polypeptide chain, is represented by sticks, with
uniform length, corresponding to the lattice spacing.
We model proteins as three-dimensional, self-avoiding chains of $N$ beads. 
To mimick amino acid interactions we use either the Miyazawa-Jernigan 
model or the G\={o} model.

\subsection{The Miyazawa-Jernigan model}

In the Miyazawa-Jernigan (MJ) model the energy of a conformation defined by 
the set of bead coordinates
$\lbrace \vec{r_{i}} \rbrace$ is given by the contact Hamiltonian
\begin{equation}
H(\lbrace \sigma_{i} \rbrace,\lbrace \vec{r_{i}} \rbrace)=\sum_{i>j}^N
\epsilon(\sigma_{i},\sigma_{j})\Delta(\vec{r_{i}}-\vec{r_{j}}),
\label{eq:no2}
\end{equation}
\noindent
where $\lbrace \sigma_{i} \rbrace$ represents an amino acid sequence,
and $\sigma_{i}$ stands for the chemical identity of bead $i$. 
The contact function $\Delta$ is $1$ if beads $i$ and $j$ are in
contact but not covalently linked and is $0$ otherwise. The interaction 
parameters $\epsilon$ are taken from the $20 \times 20$
MJ matrix, derived from the distribution of contacts of native
proteins~\cite{MJ}. \par

\subsection{The G\={o} model}

In the G\={o} ~\cite{GO} model only native contacts, i.e. contacts that 
are 
present in the native state, contribute to the energy of a conformation 
defined by $\lbrace \vec{r_{i}} \rbrace$.
In this case the contact Hamiltonian is 
\begin{equation}
H(\lbrace \vec{r_{i}} \rbrace)=\sum_{i>j}^N
B_{ij} \Delta(\vec{r_{i}}-\vec{r_{j}}),
\label{eq:no3}
\end{equation}
where the contact function $\Delta (\vec{r_{i}}-\vec{r_{j}})$, is unity only 
if beads $i$ and $j$ form a non-covalent native contact and is zero otherwise.  
Since the G\={o} model ignores the protein sequence chemical composition  
the interaction energy parameter is $B_{ij}=-\epsilon$. \par

\section{Simulation details}
Our folding simulations follow the standard MC Metropolis 
algorithm~\cite {METROPOLIS} and, in order to mimick protein movement, 
we use the kink-jump move set, including
corner flips, end and null moves as well as crankshafts~\cite{BINDER}.\par
Each MC run starts from a randomly generated unfolded conformation 
(typically with less than 10 native contacts) and the folding dynamics is 
traced by following the evolution of the fraction of native
contacts, $Q=q/Q_{max}$, where $Q_{max}$ is the total number of native 
contacts and $q$ is the number of native contacts at each 
MC step. The folding time $t$, is taken as the first passage time (FPT), that 
is, the number of MC steps corresponding to $Q=1.0$. \par
The folding dynamics is studied at the so-called optimal folding temperature, 
the temperature that minimizes the folding time as measured by the mean 
FPT.\par
The sequences studied within the context of the MJ model were prepared by 
using the design method developed by Shakhnovich and Gutin (SG)~\cite{SG} 
based on random heteropolymer theory and 
simulated annealing techniques. All targets studied are maximally compact 
structures found by homopolymer relaxation.\par

\section{Numerical results \label{sec:secno3}}
\subsection{Evidence for two folding regimes in protein folding}
Figure~\ref{figure:no1}1(a) shows the dependence on time $t$, 
of the folding probability $P_{fold}(t)$, for chain lengths 
$N=27,36,48,54,64,80,100$. Five target structures were considered per 
chain length and thirty SG sequences were prepared according 
to the method described in Ref.~\cite{SG}.\par
$P_{fold}(t)$, the probability of the chain having visited its target after 
time $t$, was computed as the fraction of (150) simulation runs, which ended 
at time $t$. 
Two distinct folding regimes were identified depending on the chain 
length. We name the regime observed for $N<80$ the {\it first regime} 
while that corresponding to $N \ge 80$ is the {\it second regime}. We have 
investigated the contribution of each target to the folding probability curve 
and found that for $N \ge 80$ the folding performance is sensitive to target 
conformation, with some targets being more foldable than others as shown in 
Figure~\ref{figure:no1}1(b) for N=100. For $N<80$ targets are equally foldable 
since all folding probability curves are consistent with asymptotic values 
$P_{fold}\rightarrow 1$.\par
In order to investigate if kinetic relaxation in the first regime is well 
described by a single exponential law we have calculated the dependence of 
$\ln 
(1-P_{fold})$ on the time coordinate $t$. Remember that in a two-state 
process the reactant concentration (the equivalent in our simulations to 
the fraction of unfolded chains) is proportional to $\exp^{-t/\tau}$ where 
$\tau$ is the so-called relaxation time. Therefore, if first regime kinetics 
is single exponential $\ln (1-P_{fold})$ $\it vs.$ $t$ should be a straight 
line with slope=$-1/\tau$. Results reported in 
Figure~\ref{figure:no2}2 show that single-exponential folding is indeed a 
very good approximation for the folding kinetics of small lattice-polymer 
proteins.

\subsection{Contact order and the lattice-polymer model kinetics}

In a recent study~\cite{PFN2} 
we analysed the folding kinetics of $\approx 5000$ SG sequences and $100$ 
target conformations distributed over the chain lengths $N=36,48,54,64$ and 
$80$.  
Targets were selected in order to cover
the observed range of CO ($\approx 0.12<$ CO $<0.26$). Results reported in 
Ref.~\cite{PFN2} show a significant correlation ($r=0.70-0.79$) between 
increasing CO and longer logarithmic folding times. In Ref.~\cite{JEWETT} 
Jewett {\it {et al.}} found a similar corelation
($r=0.75$) for a 27-mer lattice polymer modeled by a modified G\={o}-type 
potential. 
In a recent study, Kaya and Chan~\cite{KAYA2}
studied a modified G\={o} type model, with specific many-body interactions, and
found folding rates that are very well correlated ($r=0.91$) with
the CO and span a range that is two orders of magnitude larger than that of the 
corresponding G\={o} models with additive contact energies. These 
results support the empirical relation found between the contact order and the 
kinetics of two-state folders.\par

\subsection{Contact order and structural changes towards the native fold in 
the Miyazawa-Jernigan model}
In order to investigate if native geometry as measured by the CO promotes, or 
does not promote,
different folding processes, eventually leading to different folding rates, we 
have
analysed the dynamics of 900 SG sequences with chain length $N=48$, distributed
over nine target structures with low (0.126, 0.127, 0.135), intermediate 
(0.163, 0.173, 0.189) and high (0.241, 0.254, 0.259) contact order. The 
averaged trained sequence energy shows very little dispersion ranging from 
-25.11$\pm$0.03 to -26.16$\pm$0.02. Within this target set the folding time 
and the contact order correlate well ($r=0.82$) although the dispersion of 
folding times is small as reported in Figure~\ref{figure:no3}3.\par
The contact map is a $N \times N$ matrix with entries $C_{ij}=1$ if beads 
$i$ 
and $j$ are in contact (but not covalently linked) and are zero otherwise. 
Figure~\ref{figure:no4}4 shows the contact maps of targets T1 (CO=0.126), 
T2 (CO=0.189) and T3 (CO=0.259) respectively. 
One could argue that high-CO targets are associated with longer logarithmic 
folding times because they have predominantly LR contacts which, given the 
local nature of the move set used to simulate protein movement, eventually 
take a longer time to form. Let the contact time $t_{0}$ be the mean
FPT of a given contact averaged over 100 MC runs. The longest contact time 
($\ln t_{0}=12.24$) observed for target T3 is two orders of magnitude 
shorter than T3's folding time ($\ln t=17.59$)
and the sum of all contact times is $\ln(\sum_{i=1}^{57}t_{0}^{i})=15.51$, 
much lower than the
observed folding time. Thus, the fact that T3 and other high-CO 
structures have
predominantly LR contacts cannot justify, {\it per se}, their higher folding 
times.\par
The contact map provides a straightforward way to compute the frequency 
$\omega_{ij}=t_{ij}/t$ with which a native contact occurs in a MC run, 
$t_{ij}$ being the number of MC steps corresponding to $C_{ij}=1$ and $t$ 
the folding time. For each target studied we computed the mean frequency 
of each 
native contact $<\omega_{ij}>$ averaged over 100 simulation runs, and 
re-averaged $<\omega_{ij}>$ over the number of native contacts in each 
interval of backbone distance (we measure backbone distance in units of 
backbone spacing). 
We have found that while for the low-CO targets the backbone frequency 
decreases monotonically with increasing backbone distance, for the 
intermediate and high-CO targets such dependence is clearly
nonmonotonic. Figure~\ref{figure:no5}5 illustrates this behaviour for model 
structures T1, T2 and T3 elements of the low, intermediate and 
high-CO target sets respectively.    
A possible explanation for this behavior, that we have ruled out, is that of 
a negative correlation between the frequency and the energy of a contact; 
Could the most stable contacts be the most frequent ones? We found modest 
correlation coefficients $r=0.63$ and $r=0.65$ for targets T1 and T3 
respectively and therefore we conclude that the observed behaviour is not 
energy driven. \par
In Table I we show the dependence of the contact time, 
averaged over contacts in each interval of backbone distance, on the backbone 
distance for model targets T1 and T3. Since the average contact 
times, over a given range, are similar for these extreme model structures, the 
differences in the frequencies reported in Figure~\ref{figure:no5}5 must 
necessarilly distinguish different cooperative behaviors.\par
Results outlined above suggest that two broad classes of folding mechanisms 
exist for small MJ
lattice polymer protein chains. What distinguishes these two classes is the 
presence, or absence, of a monotonic decrease of contact frequency with 
increasing contact range that is related to different types of cooperative 
behaviour. The monotonic decrease of contact frequency with increasing 
backbone distance is a specific trait of low-CO structures. In this case 
folding is also less cooperative and is driven by backbone distance: Local 
contacts form first while LR contacts form progressively later as contact 
range increases.

\subsection{Contact order, long-range contacts and protein folding kinetics 
in the G\={o} model}
The energy landscapes of G\={o}-type polymers are considerably smooth because 
in the G\={o}
model the only favourable interactions are those present in the native state. 
Therefore such models are adequate for investigating the dependence of 
protein folding kinetics on target geometry.\par
In this section we investigate the contribution of LR and local interactions 
to the folding kinetics of targets T1, T2 and T3 
(Figure~\ref{figure:no4}4) in the following way: The total energy
of the native structure is kept constant but the relative contributions of LR 
and local
interactions are varied over a broad range. With the above costraint the 
energy of a conformation is given by
\begin{equation}
H(\lbrace \vec{r_{i}} \rbrace, \sigma)=C_{LR}(\sigma) H_{LR}(\lbrace \vec{r_{i}} \rbrace)+
C_{L}(\sigma)H_{L}(\lbrace \vec{r_{i}} \rbrace),
\label{eq:no4}
\end{equation}
where $C_{LR}({\sigma})={\sigma}/\lbrack(1-\sigma)Q_{L}+\sigma(1-Q_{L})\rbrack$
 and
$C_{L}({\sigma})=(1-{\sigma}) / \lbrack (1-\sigma)Q_{L}+\sigma(1-Q_{L}) \rbrack$; 
$Q_{L}$ is the fraction of local native contacts and $H_{LR}(L)$ is given by
equation~\ref{eq:no3}. The parameter $\sigma$ varies from 0 (only local 
contacts contribute
to the total energy) to 1 (only LR contacts contribute to the total energy).\par
The constraint of fixed native state energy is enforced to rule out 
differences in the folding dynamics driven by the stability of the native 
state. 
\par 
Preliminary results reported in Figure~\ref{figure:no6}6 show the dependence of the 
logarithmic folding time, 
averaged over 100 simulations runs,
on the parameter $\sigma$ for the three native geometries. 
For $\sigma < 0.15$ we have not observed folding of the target T3 and 
no 
folding was
observed for the target T2 if $\sigma < 0.10$.\par
The behaviour exhibited by target T3 is easily explained: since 
approximately 
80 percent of T3's native contacts are LR there is little competition 
between LR and local contacts.
Moreover, such competition is not significantly enhanced when one varies 
$\sigma$ towards unity. However, the effect of decreasing $\sigma$ is 
equivalent to that of `switching off' the LR contacts, that is, to force a 
structure to fold with only approximatly 20 per cent of its total native 
contacts resulting in longer folding times and for $\sigma < 0.15$ 
folding failure.       
More intriguing are the results obtained for the low and intermediate-CO 
target structures, T1 and T2 respectively. The curves 
are qualitatively similar (with a minimum at
$\sigma >0.5$) but closer inspection reveals an important difference, 
namely: for $\sigma < 0.5$ the dependence of the folding time on $\sigma$ is 
much stronger for the intermediate-CO target, T2. Indeed, in this 
case one 
observes a remarkable three-order of magnitude dispersion of logarithmic 
folding times, ranging from $\log (t)=5.62$ (for $\sigma=0.65$) to 
$\log(t)=8.50$ (for $\sigma=0.10$). 
We stress, however, that for both targets the kinetics is more sensitive 
(in the sense that the folding rate decreases more rapidly) to 
lowering $\sigma$: LR contacts appear therefore
to have a crucial/vital role, by comparison with local contacts, in 
determining the folding rates of small G\={o}-type lattice polymers and this 
effect depends on target geometry.

\section{Conclusions and final remarks}
By using different target structures in MC simulations of protein folding we 
have identified
two distinct folding regimes depending on the chain length. In close agreement 
with experimental
observations we found a first regime that describes well the folding of 
small protein molecules and whose kinetics is single exponential. Folding 
of protein chains with more than 80 amino acids, on the other hand,
belongs to a dynamical regime that appears to be target sensitive with 
some targets being
more highly foldable than others. In this case we ascribe folding failure to 
existing 
kinetic traps but we have not been able to carry out our simulations for 
long enough times in order to observe escape and succesfull folding. \par
Because the additive MJ lattice polymer model fails to exhibit the remarkable 
dispersion of folding 
rates observed in real proteins one should interpret the results 
for
the dependence of
folding times on contact order parameter with caution. However, our 
results strongly suggest that
the geometry driven cooperativity is rather robust and this 
implies an increase in folding times for increasing cooperativity.\par
We have analysed the role of LR contacts in the folding kinetics of small 
G\={o}-type lattice polymers and found a considerably strong dependence on 
target geometry. In particular, we have found that targets with a similar 
fraction of LR contacts (that is, targets with similar LRO parameter) and 
different contact order exhibit considerably different folding rates when 
LR contacts
are destabilized energetically with respect to local contacts. We are 
currently 
investigating this issue and results will be published elsewhere (in preparation). 
This result may provide a clue to
understanding the increadible dispersion of folding rates exhibited by 
real two-state folders:  
one can expect to observe longer folding times if
the distribution of contact energies in real proteins is such that local 
contacts are, on average,
more stable than LR contacts for specific native folds.

\newpage

\section{Acknowledgements}
P.F.N.F. would like to thank Funda\c c\~ao para a Ci\^encia e Tecnologia
for financial support through grant No. BPD10083/2002.

\clearpage

{\bf{FIGURES}}

\clearpage

\begin{figure}
\centerline{\rotatebox{270}{\resizebox{10.6cm}{10.6cm}
{\includegraphics{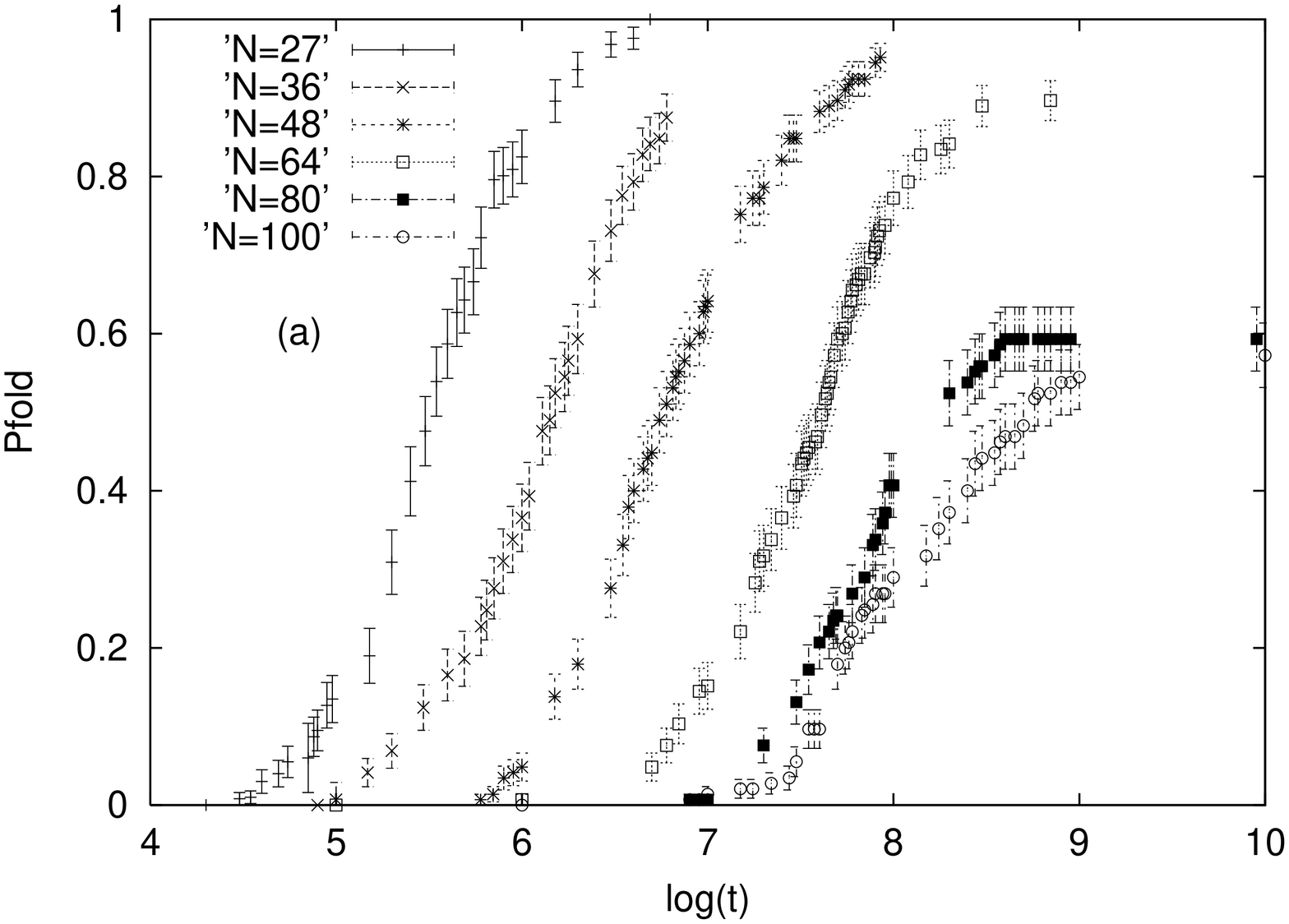}}}}
\vspace{6cm}
{\bf {Figure 1(a); P.N.F. Faisca, Biophysical Chemistry}}
\end{figure}
\clearpage

\begin{figure}
\centerline{\rotatebox{270}{\resizebox{10.6cm}{10.6cm}{\includegraphics{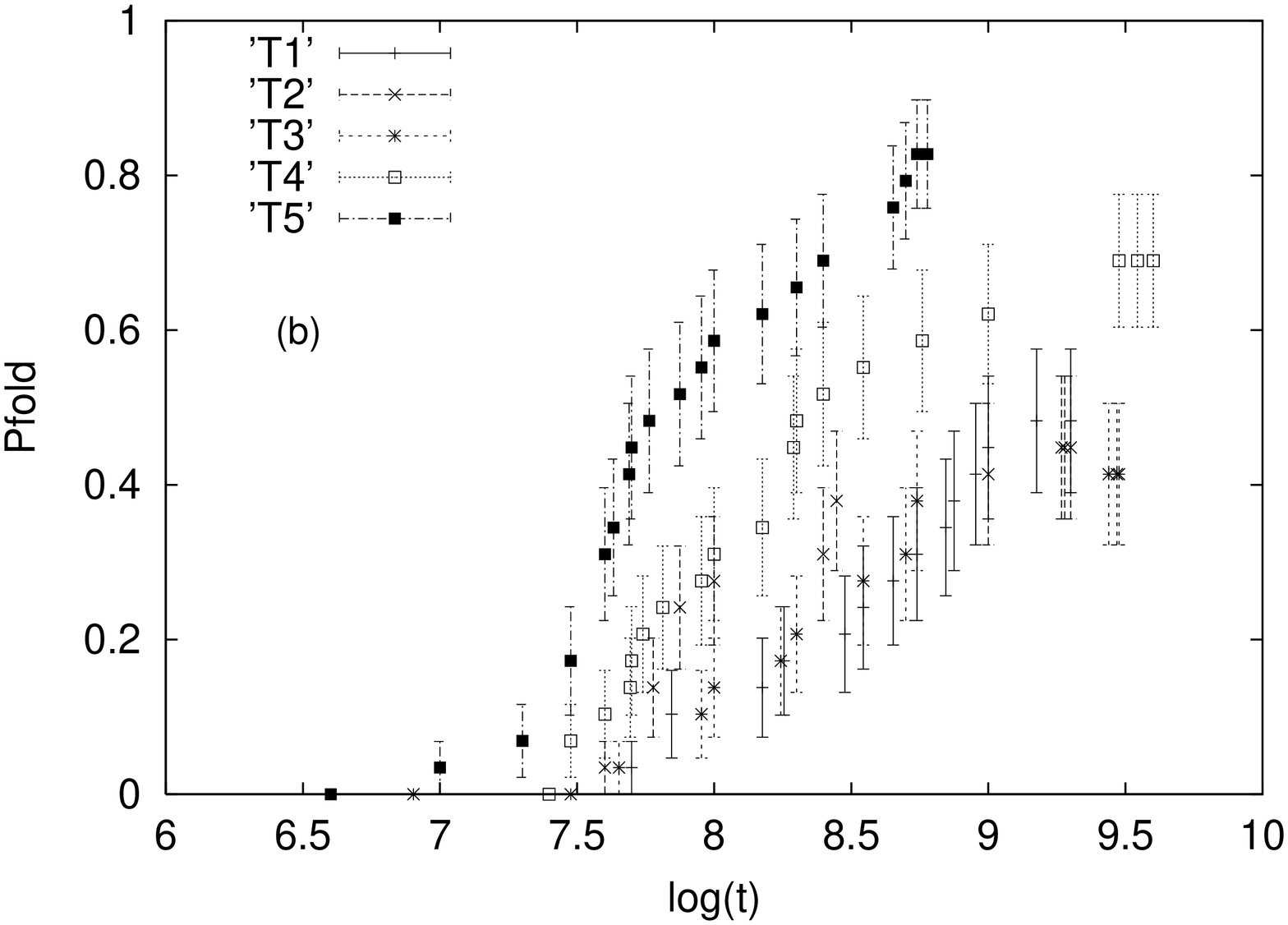}}}}
\label{figure:no1}
\vspace{8cm}
{\bf {Figure 1(b); P.N.F. Faisca, Biophysical Chemistry}}
\end{figure}
\clearpage

\begin{figure}
\centerline{\rotatebox{270}{\resizebox{12.6cm}{18.6cm}{\includegraphics{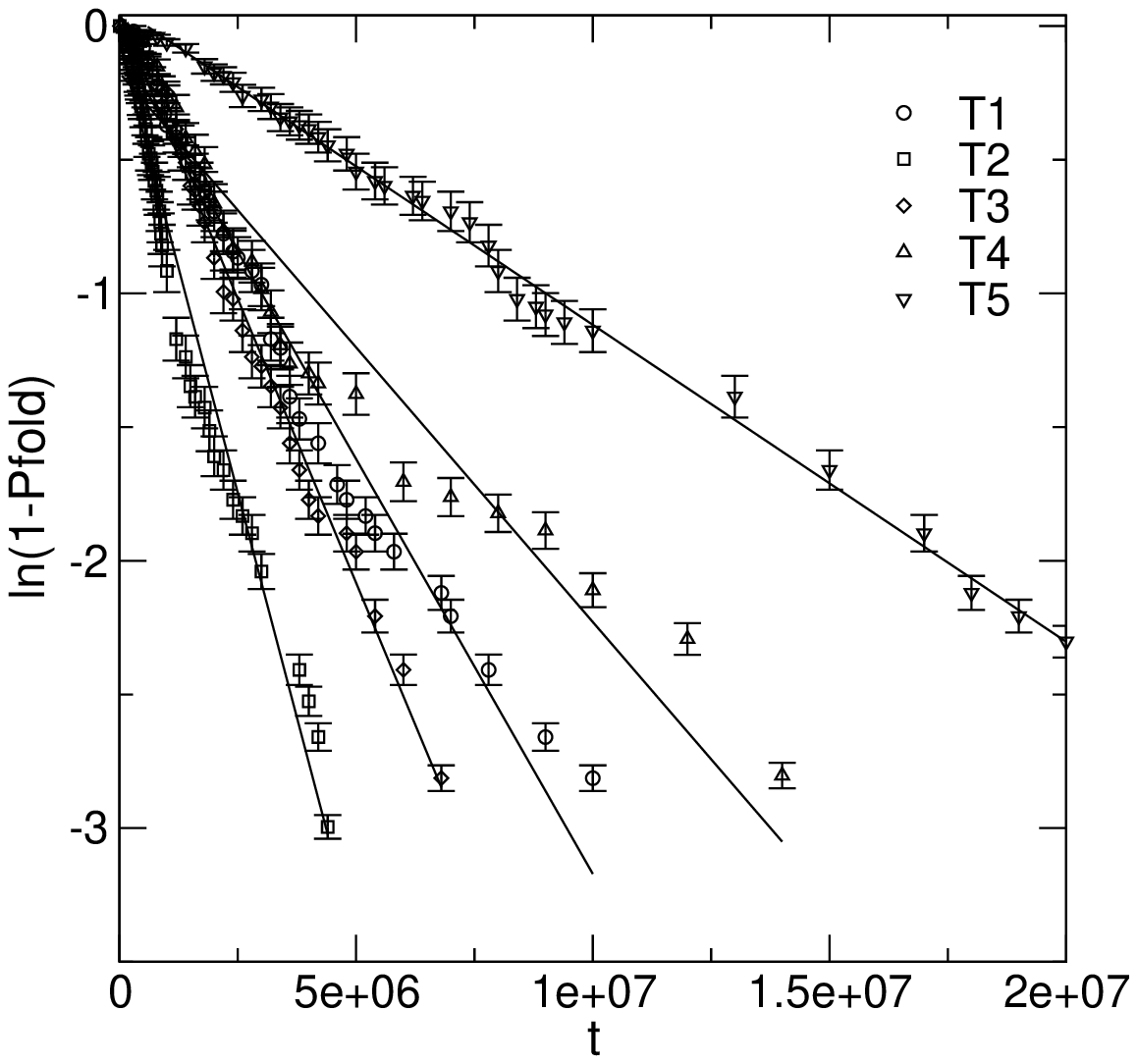}}}}
\label{figure:no2}
\vspace{8cm}
{\bf {Figure 2; P.N.F. Faisca, Biophysical Chemistry}}
\end{figure}
\clearpage

\begin{figure}
\centerline{\rotatebox{270}{\resizebox{10.6cm}{16.6cm}{\includegraphics{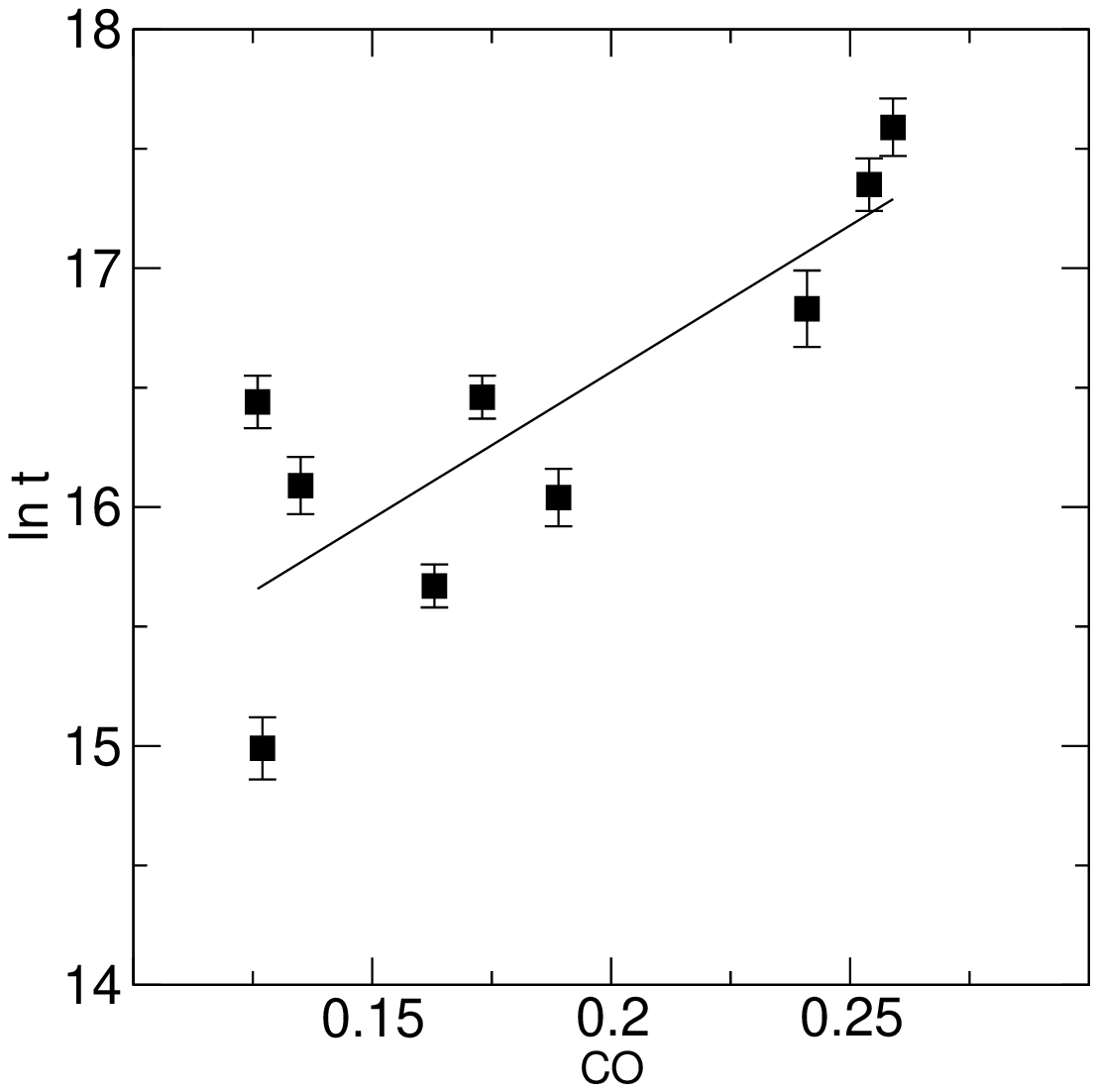}}}}
\label{figure:no3}
\vspace{8cm}
{\bf {Figure 3; P.N.F. Faisca, Biophysical Chemistry}}
\end{figure}
\clearpage

\begin{figure}
\centerline{\rotatebox{270}{\resizebox{10.6cm}{14.6cm}
{\includegraphics{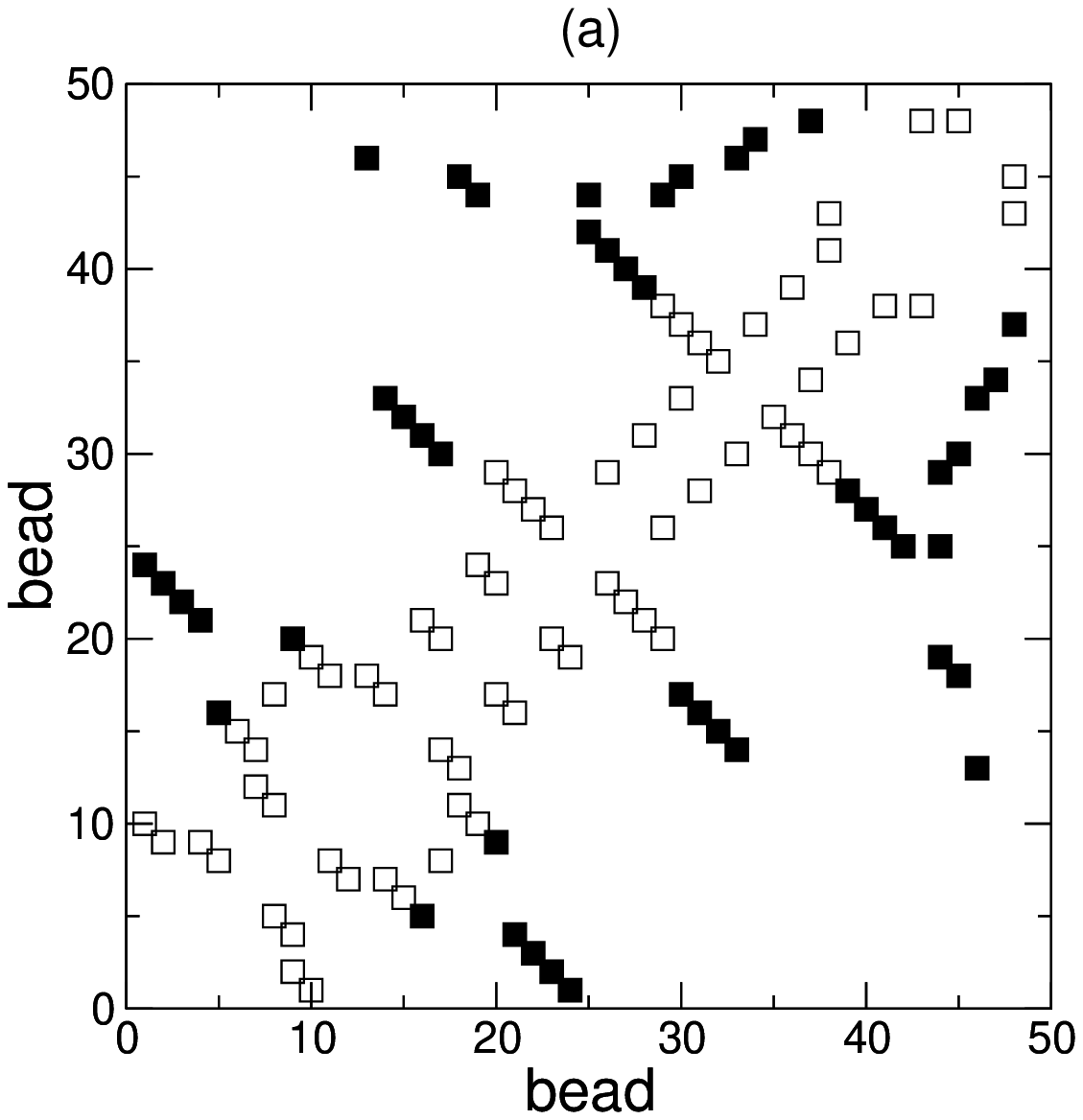}}}}
\vspace{8cm}
{\bf {Figure 4(a); P.N.F. Faisca, Biophysical Chemistry}}
\end{figure}
\clearpage

\begin{figure}
\centerline{\rotatebox{270}{\resizebox{10.6cm}{14.6cm}{\includegraphics{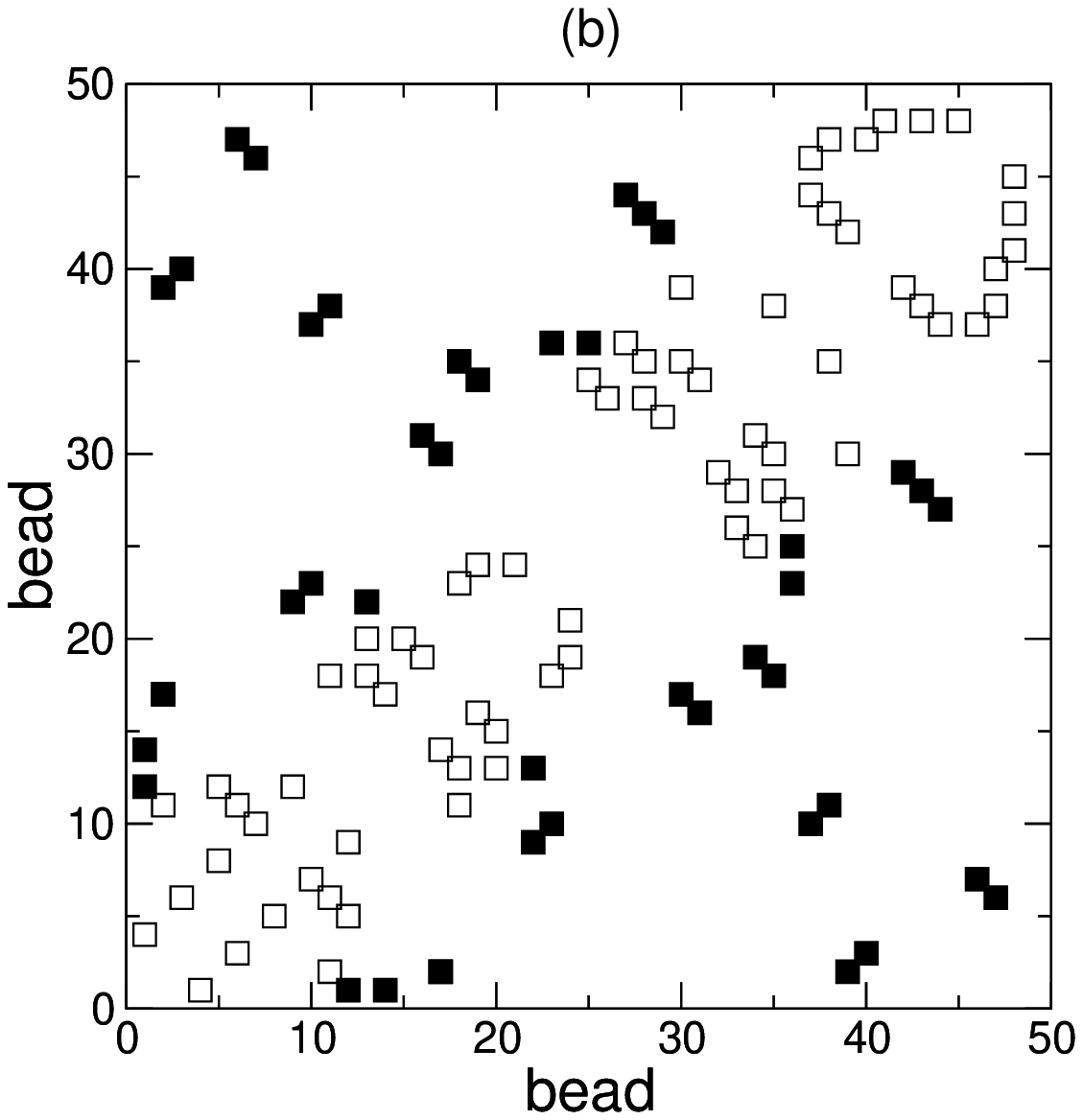}}}}
\vspace{8cm}
{\bf {Figure 4(b); P.N.F. Faisca, Biophysical Chemistry}}
\end{figure}
\clearpage

\begin{figure}
\centerline{\rotatebox{270}{\resizebox{10.6cm}{14.6cm}{\includegraphics{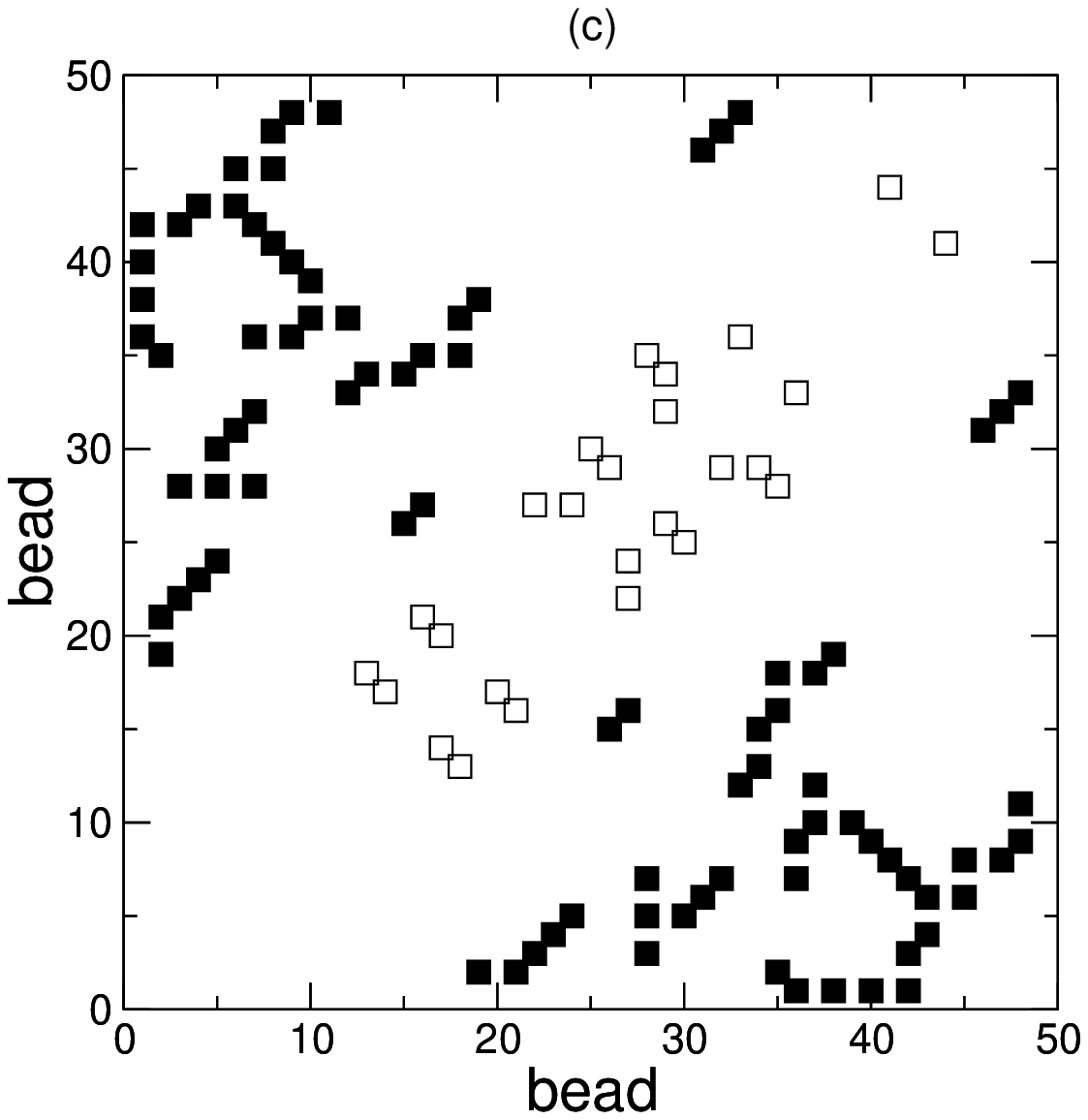}}}}
\vspace{8cm}
{\bf {Figure 4(c); P.N.F. Faisca, Biophysical Chemistry}}
\label{figure:no4}
\end{figure}
\clearpage

\clearpage

\begin{figure}
\centerline{\rotatebox{270}{\resizebox{10.6cm}{12.6cm}{\includegraphics{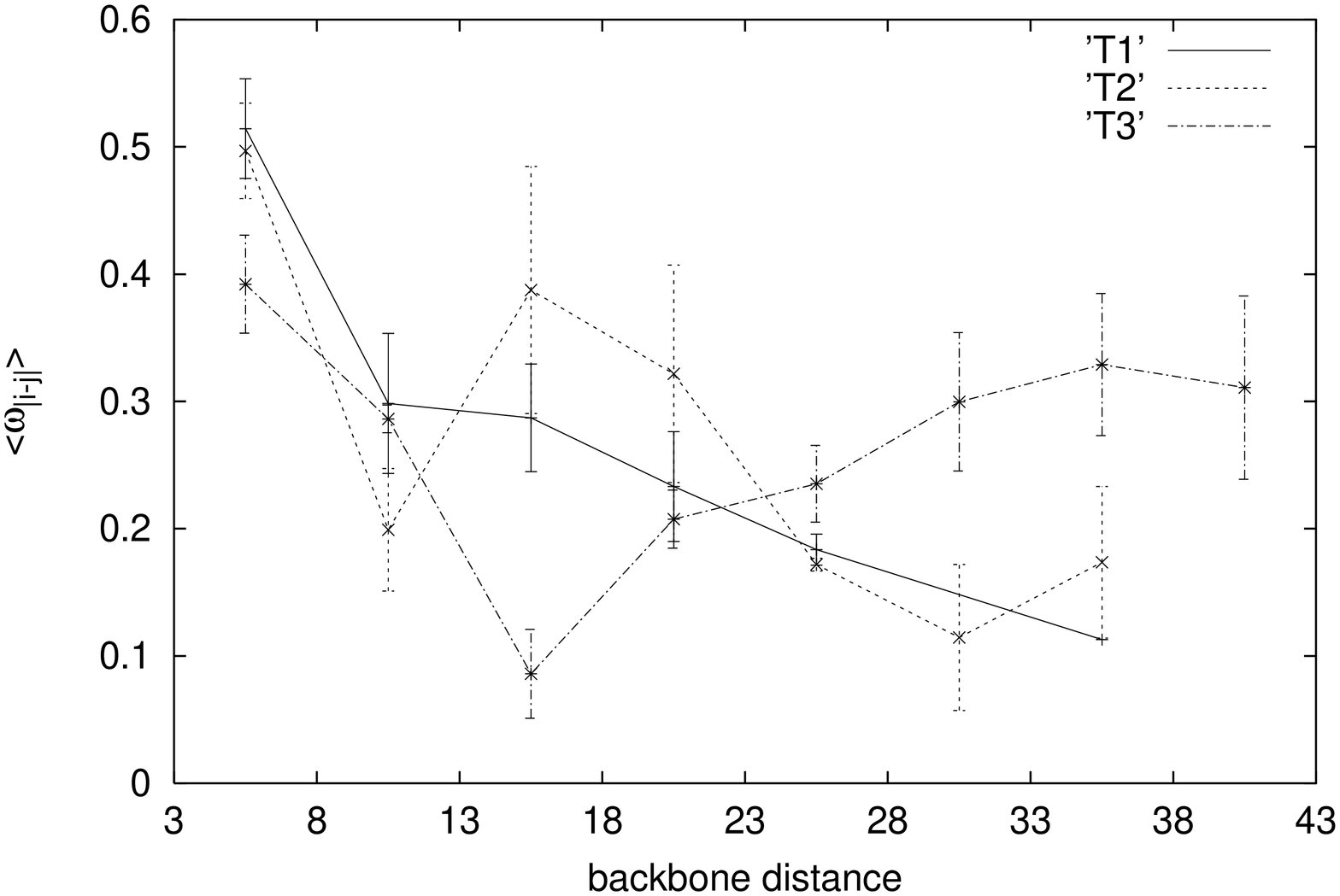}}}}
\vspace{8cm}
{\bf {Figure 5; P.N.F. Faisca, Biophysical Chemistry}}
\label{figure:no5}
\end{figure}

\clearpage

\begin{figure}
\centerline{\rotatebox{270}{\resizebox{14.6cm}{20.6cm}{\includegraphics{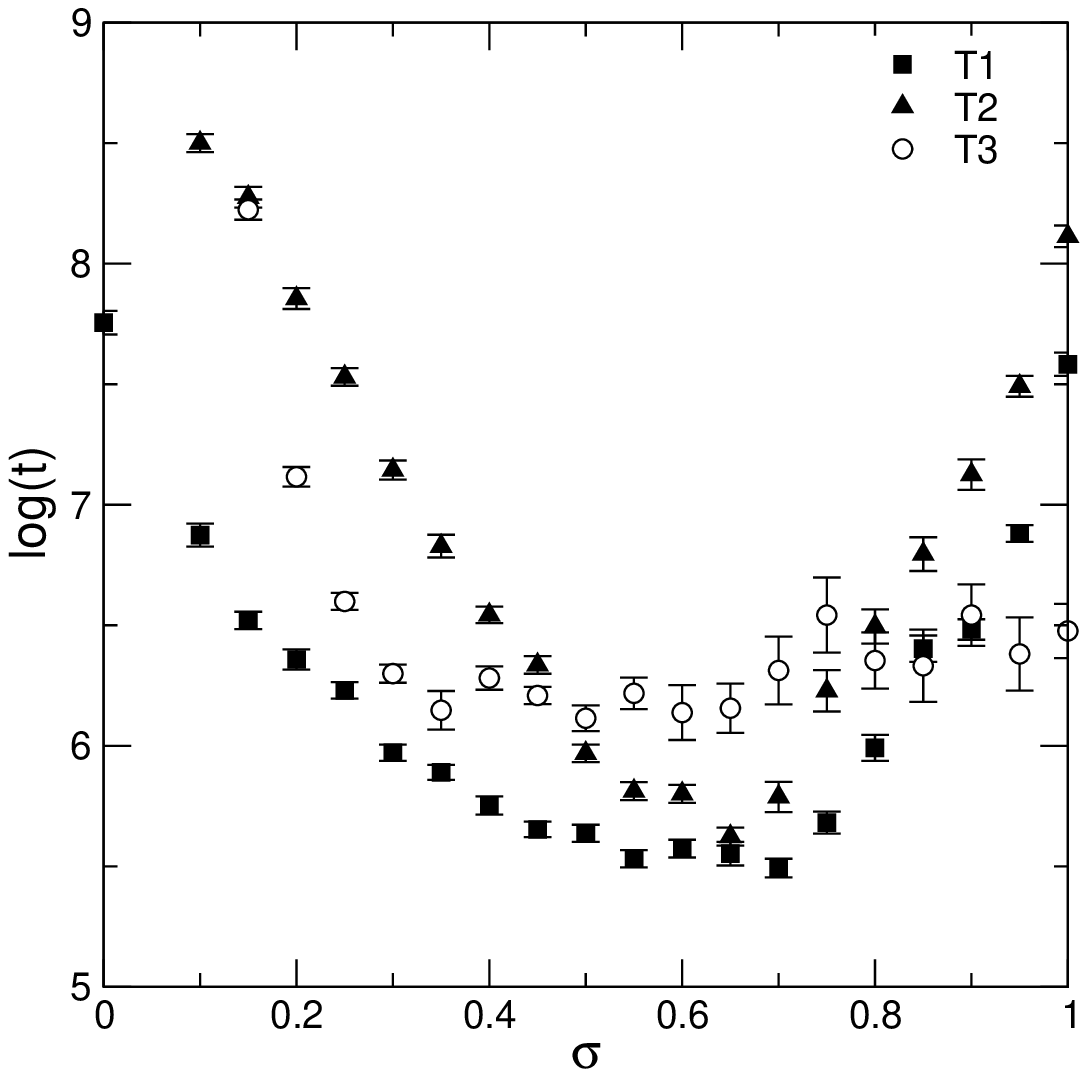}}}}
\vspace{8cm}
{\bf {Figure 6; P.N.F. Faisca, Biophysical Chemistry}}
\label{figure:no6}
\end{figure}

\clearpage

{\bf{FIGURE CAPTIONS}}

\clearpage

Figure 1. Dependence of the folding probability, $P_{fold}$, on $\log(t)$. (a)
For each of the chain lengths $N=27,36,48,64,80$ and $100$ five target 
structures were considered and 30 sequences were designed per target. $P_{fold}$, the probability of the chain having visited its target after time $t$, was computed as the fraction of 
simulation runs that ended in time $t$. (b) Separate contribution of 
each of the 100 bead long targets for the dependence of $P_{fold}(t)$ on $\log(t)$ ~\cite{FAISCA}.
\vspace{1cm}

Figure 2. Evidence for single exponential folding kinetics for chain length 
$N=48$. The correlation coeficient between the logarithmic fraction of unfolded chains 
and `reaction' time is $r \approx 0.97$ for target T4 and $r \approx 0.99$ for the remaining targets.
\vspace{1cm}

Figure 3. Dependence of the logarithmic folding times, $\ln_{e} t$,  on the contact order 
parameter ($r \approx 0.82$).
\vspace{1cm}

Figure 4. {Contact maps of targets T1 (a), T2 (b) and T3 (c). Each square 
represents a native contact. We divide the 57 native contacts into two classes: LR 
contacts are represented by filled squares and correspond to contacts between 
beads for which the backbone separation is 10 or more backbone units. Local 
contacts are represented by white squares. There are 23 LR contacts
in structure T1, 21 in structure T2 and 44 in structure T3.}
\vspace{1cm}

Figure 5. The backbone frequency, $<\omega_{\vert i-j \vert}>$, as a function 
of the backbone separation for the low-CO, high-CO and intermediate-CO target. 
The backbone frequency is the mean value of $<\omega>$ averaged over the 
number of contacts in each interval of backbone separation.
\vspace{1cm}

\newpage

Figure 6. Dependence of the logarithmic folding time $\log_{10} t$ on the 
parameter sigma. The parameter $\sigma$ varies from 0 (only local
contacts contribute to the total energy) to 1 (only LR contacts contribute to the total energy).

\clearpage

{\bf{TABLES}}

\clearpage

\begingroup
\squeezetable
\begin{table}
\caption{The averaged contact time, $\ln_{e}<t_{0}>$, as a function of the
backbone separation ~\cite{FAISCAPRE}}\label{tab:tabno1}
\begin{ruledtabular}
\begin{tabular}{lcccccccc} 
Target & \multicolumn{8}{c}{backbone distance} \\ \cline{2-9}
& \multicolumn{1}{c}{$ \lbrack 3, 8 \lbrack $}
& \multicolumn{1}{c}{$ \lbrack 8, 13 \lbrack $}
& \multicolumn{1}{c}{$ \lbrack 13, 18 \lbrack$}
& \multicolumn{1}{c}{$ \lbrack 18, 23 \lbrack$}
& \multicolumn{1}{c}{$ \lbrack 23, 28 \lbrack$}
& \multicolumn{1}{c}{$ \lbrack 28, 33 \lbrack$}
& \multicolumn{1}{c}{$ \lbrack 33, 38 \lbrack$}
& \multicolumn{1}{c}{$ \lbrack 38, 43 \lbrack$}
\\\hline
$T1$ & 8.14 $\pm$ 0.12 & 10.67 $\pm$ 0.21 & 11.47 $\pm$ 0.07 & 11.69 $\pm$ 0.13 & 11.39 $\pm$ 0.07 & - &
11.58 $\pm$ 0.10  & -   \\ 
$T3$ & 7.66 $\pm$ 0.10 & 11.14 $\pm$ 0.18 & 10.88 $\pm$ 0.10 & 11.60 $\pm$ 0.06 & 12.06 $\pm$ 0.05 &
12.24 $\pm$ 0.08 & 11.82 $\pm$ 0.05 & 11.61 $\pm$ 0.05 
\end{tabular}
\end{ruledtabular}
\end{table} 
\endgroup
\clearpage

\end{document}